\documentclass[preprint,amsmath,nofootinbib,twoside]{revtex4}
\usepackage{graphicx}

\newcommand{\eqn}[1]{\label{eq:#1}}
\newcommand{\refeq}[1]{(\ref{eq:#1})}
\newcommand{\eq}{eq.~\refeq}

\newcommand{\reffig}[1]{(\ref{fig:#1})}
\newcommand{\fig}{fig.~\reffig}

\begin{document}

\title{A hidden solution to the $\mu/B_\mu$ problem in gauge mediation}
\date{\today}

\author{Tuhin S. Roy}
\email{tuhin@bu.edu}
\author{Martin Schmaltz}
\email{schmaltz@bu.edu}
\affiliation{Physics Department,  Boston University\\
  Boston, MA 02215, USA}
\preprint{BUHEP-07-06}
\begin{abstract}
  We propose a solution to the $\mu/B_\mu$ problem in gauge mediation.
  The novel feature of our solution is that it uses dynamics of the hidden 
  sector, which is often present in models with dynamical supersymmetry 
  breaking. We give an explicit example model of gauge mediation where 
  a very simple messenger sector generates both $\mu$ and $B_\mu$ 
  at one loop. The usual problem, that   $B_\mu$ is then too large, is
  solved by strong renormalization effects from the hidden sector which
  suppress $B_\mu$ relative to $\mu$. Our mechanism relies on an
  assumption about the signs of certain incalculable anomalous dimensions
  in the hidden sector. Making these assumptions not only allows us
  to solve the $\mu/B_\mu$ problem but also leads to a characteristic
  superpartner spectrum which would be a smoking gun signal for our
  mechanism.
\end{abstract}

\maketitle

Models with  Gauge Mediated Supersymmetry Breaking 
\cite{Dine:1981za,Dimopoulos:1981au,Dine:1981gu,Nappi:1982hm,
AlvarezGaume:1981wy,Dimopoulos:1982gm,Dine:1993yw,Dine:1994vc,Dine:1995ag,
Giudice:1998bp} are attractive because they introduce no
new flavor violation beyond the Standard Model.%
\footnote{for recent developments see for example \cite{Dine:2006xt,
Kitano:2006xg,Murayama:2006yf,Csaki:2006wi,Aharony:2006my,Murayama:2007fe,
Delgado:2007rz,Abel:2007jx}
}
 However, gauge mediation 
is not free of problems. In this paper we are concerned with the 
$\mu/B_\mu$-problem \cite{Dvali:1996cu},
which is particularly severe in gauge mediation. Solutions to the 
$\mu/B_\mu$-problem usually involve an elaborate messenger sector or extra 
light particles, often requiring fine-tuning of parameters.
In this paper we point out an alternative solution to the 
$\mu/B_\mu$-problem, which does not require a complicated
messenger sector or new particles at the weak scale.

The $B_\mu$-problem in gauge mediation is related to the 
$\mu$-problem, which is common to all supersymmetric models.  
The effective low energy MSSM Lagrangian contains a supersymmetric Higgs 
mass term
\begin{equation}
  \int\! d^2\theta\, \mu H_u H_d \; .
 \label{eq:l-mu}
\end{equation}
Natural electroweak symmetry breaking requires that the mass parameter
$\mu$ is of the same size as superpartner masses. Relating the 
supersymmetry preserving
$\mu$ parameter to the supersymmetry violating soft masses 
is the $\mu$-problem in supersymmetric theories.
In supergravity, Giudice and Masiero proposed a simple solution
\cite{Giudice:1988yz}:
the $\mu$-term stems from a higher dimensional 
operator coupling the 
supersymmetry breaking field $X$ to the MSSM Higgs fields, such that the 
$\mu$-term in \eq{l-mu} is generated when $X$ is replaced by its 
supersymmetry breaking vacuum expectation value $F_X$.  
\begin{equation}
 \label{eq:mu-op}
   \int\! d^4\theta\, k_{\mu} \frac{1}{M} X^\dagger H_u H_d \quad 
             \rightarrow 
    \quad \int\! d^2\theta\, \mu H_u H_d  \quad \text{with} \quad 
      \mu = k_{\mu} \frac{F_X}{M} \; .
\end{equation}
Here $M$ is the mediation scale, it is given by $M_{\rm Planck}$ in 
supergravity theories. The gaugino masses $M_a$ ($a$ labels the three SM 
gauge groups) come from similar higher dimensional operators 
\begin{equation}
   \label{eq:ino-op} 
  \int\! d^2\theta\, w_a \frac{1}{M} X W^a W^a  \quad 
             \rightarrow 
      \quad   M_a = w_a \frac{F_X}{M} \; .
\end{equation}
If all these operators are generated by supergravity we may assume that
the coupling constants $k_{\mu}$ and $w_a$ are of order one and we 
find $\mu \sim M_a$ as desired.

Another important part of the Higgs potential is the $B_{\mu}$-term.
Natural electroweak symmetry breaking requires
$B_{\mu} \sim  \mu^2$. The $B_{\mu}$-term arises from a higher dimensional 
operator (the $B_{\mu}$-operator)
\begin{equation}
  \label{eq:bmu-op} 
   \int\! d^4\theta\, k_{B}\frac{1}{M^2}   X^\dagger X H_u H_d \quad 
             \rightarrow \quad 
     B_{\mu} =  k_{B}  \frac{\left|F_X\right|^2}{M^2}  \; .
\end{equation}
In many models, $k_B$ is generated with a similar size as  $k_{\mu}$ and 
$w_a$. If, in addition, $k_B \sim k_\mu \sim w_a \sim \mathcal{O}(1)$ as
in minimal supergravity with the Giudice-Masiero mechanism, then we obtain
the desired relations  $B_{\mu} \sim  \mu^2 \sim M_a^2$.

\begin{figure}[t]
\begin{center}
\includegraphics[width=\textwidth]{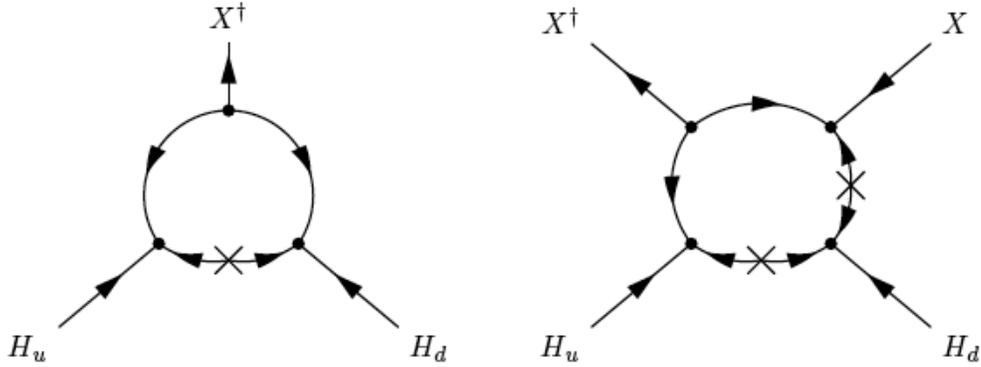}
\end{center}
\caption{Superfield diagrams which generate the $\mu$ and 
 $B_{\mu}$-operators at one loop. The fields in the loop are messengers.
 A specific superpotential with couplings which generate such diagrams 
 is given in \eq{w-mess}.}
\label{fig:mu-gen}
\end{figure}

The situation is more complicated in models with gauge mediation.
Here gaugino and scalar masses are generated from gauge loop diagrams
involving messenger fields of mass $M$. The $\mu$ and $B_\mu$ terms
cannot be generated by gauge loops because the operators
eqs.~[\ref{eq:mu-op},\ref{eq:bmu-op}] are forbidden by a Peccei-Quinn symmetry.
A simple way to generate a $\mu$-term is to break Peccei-Quinn symmetry
by coupling the  Higgs superfields to the messengers in the superpotential.
Now $\mu$ and $B_\mu$ terms are generated by the one loop diagrams
shown in \fig{mu-gen} and we expect $k_B \sim k_\mu \sim w_a \sim 1/16 \pi^2$.
Hence  $\mu$ is naturally of the size of the gaugino masses which also
arise at one loop.
However, the fact that the $B_{\mu}$ operator is also generated
at one loop implies that the mass-squared  parameter $B_\mu$ is too
large by a loop factor compared to $\mu^2$ and the gaugino masses squared
\begin{equation}
  B_{\mu} \sim 16 \pi^2 \mu^2 \sim  16 \pi^2  M_a^2 \; .
\end{equation}
This is the $B_\mu$ problem in gauge mediation.

Solutions to the above problem are non-trivial. The basic point of the 
solution in \cite{Dvali:1996cu} 
is to design the messenger superpotential such 
that at the leading order only the $\mu$-term is generated - the $B_\mu$
operator in \eq{bmu-op} is generated at higher order. 
Models based on this scheme require extra heavy gauge singlets with carefully 
chosen masses and couplings. Another 
popular scheme is to introduce light scalars as in the 
NMSSM. However, since it is difficult to obtain soft masses for
gauge singlets in gauge mediation, one usually ends up either
fine-tuning electroweak symmetry breaking or predicting
unacceptably light particles.
For a review of the $B_\mu$ problem in gauge mediation
and references see \cite{Giudice:1998bp}. 

In this paper we propose a new solution to the $B_{\mu}$-problem: we
start with a very simple messenger sector,
requiring that it generates a $\mu$-term of the right 
size, but allowing that the $B_{\mu}$ term at the messenger scale  is
too large. We then argue that renormalization effects due to strong hidden sector 
interactions can sufficiently suppress the  $B_{\mu}$-operator relative to the 
$\mu$-operator at low energies. This relative renormalization between 
$\mu$ and $B_\mu$ due to hidden
sector interactions was also pointed out in \cite{Dine:2004dv}.

Let us explain our mechanism in more detail. First, note that since
the MSSM interactions are weak they cannot significantly suppress the
$B_\mu$ term due to renormalization. This is why solutions to the $B_\mu$ 
problem generally require $B_\mu \lesssim \mu^2$ at the
messenger scale even though phenomenology only requires this
condition near the weak scale. On the other hand, in many models of
dynamical supersymmetry breaking hidden sector interactions are
strong and can induce large renormalization.
For our mechanism we require a large
positive anomalous dimension for the $B_\mu$ operator to suppress
$B_\mu$ relative to $\mu^2$ at low energies.

A formalism for renormalization which takes into account arbitrary
hidden sector interactions as well as MSSM interactions was
developed in \cite{Cohen:2006qc,crs2}. We follow this formalism and
notation, and adopt a holomorphic basis in which no wave-function
renormalization is performed. We begin with the renormalization of the 
$\mu$-operator. It is well-known that in the holomorphic basis the 
superpotential is not renormalized. This non-renormalization theorem can
be generalized to K\"ahler potential operators which factor into a 
product of a chiral visible sector operator times an anti-chiral
hidden sector operator
\begin{equation}
\frac{1}{M^d}\left. \,   {\cal O}_h^* {\cal O}_v\right|_{\theta^4} \; .
\label{eq:hidvisop}
\end{equation}
The proof assumes that the
dimension of the operator in question is low enough so that its 
renormalization can only involve single insertions of higher-dimensional
hidden-visible interactions.
Then its renormalization factorizes into separate visible and
hidden sector contributions, and we can compute these contributions
independently. To compute the visible sector running, we treat the
hidden sector fields as background fields with an expectation value
for their $\overline \theta^2$ components. This turns the operator
in \eq{hidvisop}
into a chiral superpotential for the visible fields which is 
protected by the usual non-renormalization theorem.
By supersymmetry, this is also true for the full operator
with arbitrary hidden sector fields. Similarly, as far as
purely hidden sector interactions are concerned the operator in
\eq{hidvisop} is anti-chiral and therefore not renormalized.

\begin{figure}[t]
\begin{center}
\includegraphics[width=0.55\textwidth]{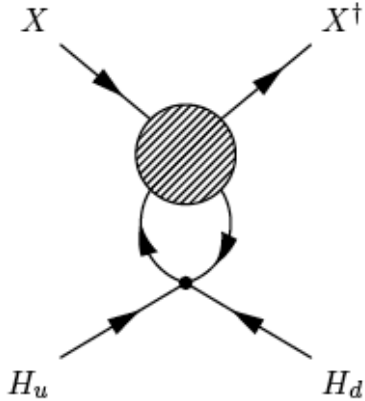}
\end{center}
\caption{Interactions of the hidden sector field $X$  renormalize 
   $X^\dagger X H_u H_d$}
\label{fig:bmu-ren}
\end{figure}

Applying this result we see that the $\mu$ operator is not renormalized. 
Using a similar argument
we can show that the $B_\mu$ operator is not renormalized
by visible sector interactions because it is chiral in visible fields.
However, hidden sectors interactions do renormalize $B_\mu$ because
the operator is real in hidden sector fields and therefore not protected by
nonrenormalization theorems. Schematically, these renormalizations
are represented by the diagram in \fig{bmu-ren}. 

This implies the following general form for the renormalization group 
equations of the couplings $k_\mu$ and $k_B$ (again in the holomorphic basis)
\begin{equation}
  \label{eq:rge}
  \begin{split}
   \frac{d}{dt} k_{\mu} = 0    \quad &\Rightarrow \quad
        k_{\mu}\big|_E  = k_{\mu}\big|_M   
     \\
    \frac{d}{dt} k_{B} =   \gamma  k_{B}  \quad &\Rightarrow \quad 
  k_{B}\big|_E  =   \exp\left(\!-\!\int_{t}^{0} \!\! ds\,
    \gamma(s)\!\right) \;   k_{B}\big|_M \equiv G(\frac{E}{M}) \;k_{B}\big|_M \;,
  \end{split}
\end{equation}
where $t=\ln(E/M)$ and $E$ is the renormalization scale. 
$\gamma$ is the anomalous dimension of the operator in the holomorphic 
basis.  For example, if  $\gamma$ is constant, then $G(E/M) = (E/M)^\gamma$.

We now see that if $\gamma$ is positive and of order one then we can
easily obtain large suppression factors $G \lesssim 1/16 \pi^2$ for the 
$B_\mu$ operator at low energies. This then implies that 
$B_\mu \lesssim \mu^2$ at the scale where the hidden sector dynamics ends.
For our mechanism to work the hidden sector 
must be strongly coupled over at least a couple of decades of energy scales.
The most familiar such theories are strongly coupled
conformal field theories.

Now we describe an explicit model which demonstrates our mechanism.
The model has the following features:
a very economical messenger sector which naturally predicts a 
$\mu$ parameter of the size of the gaugino masses (but $B_\mu$ is too 
large at the messenger scale). Below the messenger scale, our hidden sector 
fields have strong, approximately conformal interactions
which lead to large anomalous dimensions for at least a couple of decades of
running. Assuming that the anomalous dimensions governing the running of $B_\mu$
are positive we find that $B_\mu$ is suppressed to acceptably small values.
Finally, at an even lower scale, supersymmetry is broken 
spontaneously in the hidden sector.

A simple messenger sector that suffices our purposes  was described in 
\cite{Dvali:1996cu}. The messengers $R_1, R_2, \overline R_1, \overline R_2$
are vector like under the standard model  gauge group and couple to the
gauge singlet $X$ and the MSSM Higgses in the superpotential
\begin{equation}
  \eqn{w-mess}
    W_{\text{messenger}} = \bar{R}_1 \left( M + X \right) R_1
     +\bar{R}_2 \left( M + X \right) R_2 +  H_u \bar{R}_1 R_2 + 
      H_d \bar{R}_2 R_1 \; ,
\end{equation}
where all Yukawa couplings are assumed to be of order one.
At the scale $M$ the messengers are integrated out and the higher dimensional
operators for $\mu$, $B_\mu$, as well as gaugino and scalar masses are 
generated. The operators in 
eqs.~[\ref{eq:mu-op},\ref{eq:ino-op},\ref{eq:bmu-op}] all
appear at one loop which implies the relation
$w_a \big|_M \sim k_\mu \big|_M \sim k_B \big|_M \sim 1/16\pi^2$ 
({\it i.e.} the $B_\mu$ term is too large: $k_B \sim 16\pi^2 k_\mu^2$). 

As our model of the hidden sector we take supersymmetric QCD
with gauge group $SU(N)$ and $F$ flavors each of  $Q+\overline Q$ and 
$P + \overline P$ in the fundamental and anti-fundamental representations.
The supersymmetry breaking field $X$ is a gauge singlet and couples to
$Q$ and $\overline Q$ with the following hidden sector superpotential  
\begin{equation}
  \label{eq:w-hidden}
X(\Lambda^2 + \lambda Q\overline Q) + m (Q \overline P + P \overline Q)
 + \kappa (Q \overline Q)^2  \; .
\end{equation}
Here $m$ and $\Lambda$ are mass parameters with $m > \Lambda$,
$\lambda$ is a Yukawa coupling, $\kappa$ is a coupling to be 
discussed below, and all hidden sector flavor indices are contracted
with $\delta_{ij}$. For $\frac{3}{2} N < 2F < 3 N$ and at energies
above $m$ the theory approaches an infrared attractive conformal fixed point
for both gauge and Yukawa couplings \cite{Seiberg:1994pq}.
During this running the suppression of scalar masses and $B_\mu$ takes
place. At the scale $m$ conformal symmetry is broken, the
field $X$ decouples from the gauge dynamics and 
supersymmetry breaks at the scale $\Lambda$. While this story 
appears simple, there are subtleties which will require us to modify
the model. We will now discuss these subtleties.

First, we study the renormalization of operators which couple hidden and 
visible fields in the energy regime $m< E < M$ where the hidden sector is 
governed by strong conformal dynamics. As already shown above, operators 
coupling chiral hidden sector operators to visible fields are not 
renormalized in the holomorphic
basis which we employ. The gauge-invariant non-chiral operators with the 
lowest scaling dimensions of the hidden sector are presumably the bilinears 
$X^\dagger X$, $Q_i^\dagger Q_j$, $\overline Q_i^\dagger \overline Q_j$, 
$P_i^\dagger P_j$, and $\overline P_i^\dagger \overline P_j$. In general, any
of these operators  may couple to the Higgs bilinear
\begin{equation}
  \label{eq:toy-ops}
  \int\! d^4\theta\, \frac{1}{M^2}  H_u H_d \left(  k_X X^\dag X 
  + k^{ij}_{Q} Q_i^\dag Q_j + \overline k_{\overline{Q}}^{ij}  
                        \overline{Q}_i^\dag \overline{Q}_j  
  + k^{ij}_{P} P_i^\dag P_j + \overline k_{\overline{P}}^{ij}  
                        \overline{P}_i^\dag \overline{P}_j  \right) \; .
\end{equation} 
Integrating out the messengers at scale $M$ only generates $k_X$
but hidden sector interactions can generate other
couplings through operator mixing.  At one loop, 
the operator $Q^\dagger Q + \overline Q^\dagger \overline Q$ is generated
from the Yukawa interaction, and at two loops 
the operator $P^\dagger P + \overline P^\dagger \overline P$ is generated
from gauge interactions.
All other possible operators in \eq{toy-ops}
are not generated. It is easy to understand why they are not generated by 
considering the global symmetries of the model. Setting the coupling $\kappa=0$
for the moment, only the dimensionless $\lambda$ and the $SU(N)$ gauge coupling
are relevant to the renormalization. Both preserve
$SU(F)_Q \times SU(F)_P \times SU(F)_{\overline P}$
nonabelian flavor symmetries and several $U(1)$ symmetries. The non-abelian
flavor symmetries together with the $U(1)$ symmetries forbid all operators 
except
$X^\dagger X\ , Q^\dagger Q +  \overline Q^\dagger \overline Q\ , 
Q^\dagger Q -  \overline Q^\dagger \overline Q\ , 
P^\dagger P +  \overline P^\dagger \overline P\ ,
P^\dagger P -  \overline P^\dagger \overline P$.
The two operators with the minus signs contain N\"other currents 
corresponding to two baryon number symmetries acting separately on the 
$Q, \overline Q$ and $P, \overline P$. The baryon number symmetries are 
preserved by the interactions of the CFT and therefore the corresponding 
currents are conserved. Current conservation implies that these operators 
are not renormalized in the canonical basis for the fields. The 
holomorphic basis differs from the canonical one only by wave function 
renormalization. And since wave function renormalization is multiplicative
new couplings to these currents cannot be generated in the holomorphic basis either. 
The current $P^\dagger P +  \overline P^\dagger \overline P$ is 
broken by the axial anomaly but it is easy to see that diagrams 
proportional to the anomaly only enter its renormalization at two loops. 
This leaves us with three operators which mix in the running of $B_\mu$
to all orders
\begin{equation}
  \label{eq:2-ops}
  \int\! d^4\theta\, \frac{H_u H_d}{M^2}  \left(  k_X X^\dag X 
 + \frac{k_Q}{\sqrt{2NF}} (Q^\dag Q+\overline Q^\dag \overline Q)
 + \frac{k_P}{\sqrt{2NF}} (P^\dag P+\overline P^\dag \overline P)\right) \; .
\end{equation} 
The operator coefficients $k$ satisfy the renormalization group equation
\begin{equation}
  \label{eq:rge-cft}
    \frac{d}{dt} k =   \gamma\, k \; ,
\end{equation}
where $\gamma$ is a $3\times3$ dimensional matrix of anomalous dimensions.
Assuming that our theory is approximately conformal, the anomalous dimensions
are approximately constant. Diagonalizing $\gamma$
and denoting its smallest eigenvalue by $\gamma_<$,  we find
that the operators in \eq{2-ops}
are suppressed by a mixing angle times $\left(\frac{E}{M}\right)^{\gamma_<}$
at low energies $E$. The suppression turns into an
enhancement if $\gamma_<$ is negative. In order to achieve sufficient 
suppression of the $B_\mu$ operator we see that we must require
$\left(\frac{E}{M}\right)^{\gamma_<} \lesssim 1/16 \pi^2$
or $\gamma_< \gtrsim {\rm log}(16 \pi^2) / {\rm log} (M/E)$
where $E$ is now the energy scale at which the hidden sector stops 
interacting. To put it simply, all eigenvalues of the anomalous dimension 
matrix must be of order one and positive.

Does our model satisfy this criterion? In the one loop
approximation $k_P$ vanishes and the anomalous dimension matrix $\gamma$ 
is only $2\times2$ 
\begin{equation}
\gamma= \frac{1}{16 \pi^2}
\left(
\begin{array}{cc}
  0& 2 \lambda^2 \sqrt{2NF}   \\
  2\lambda^2\sqrt{2NF}& 2\lambda^2+g^2 (N^2-1)/N  
\end{array}
\right) \ .
\end{equation}
This matrix has one positive and one negative eigenvalue
which would be a problem.
At strong coupling, the anomalous dimensions are not calculable,
they are expected to be non-vanishing and of order unity
but we do not know their sign.

We will now show that for hidden sector operators corresponding to
conserved hidden sector currents of the CFT
the sign can be determined and is negative. Thus they must be
avoided in building suitable hidden sector models by
either breaking the corresponding symmetry in the CFT or by 
preserving the symmetry also in the messenger sector so that 
the dangerous operators are never generated. The argument
for why conserved current operators are dangerous is simple.
In the canonical basis (which we will adopt for the argument in this
paragraph) such a conserved current is not renormalized.
Therefore the operator coupling the current to the
MSSM fields scales as $1/M^2$ even in the presence of strong interactions.
On the other hand, the $\mu$ term arises from a gauge-singlet
anti-chiral operator of the hidden sector multiplying the MSSM Higgs fields.
This operator has a positive anomalous dimension in the canonical basis.
This follows from unitarity arguments which imply that the dimensions
of chiral gauge invariant operators must be greater than one.
Thus in the presence of operators corresponding to conserved currents of 
the CFT $B_\mu$ actually increases compared to $\mu^2$.
Therefore exact currents of the CFT which are broken by messenger physics must
be avoided for our mechanism to work.%
\footnote{This is similar to the constraints
on currents of hidden sector CFTs for conformal sequestering
\cite{Schmaltz:2006qs}.}
Equivalently, in the holomorphic basis, the $\mu$-operator is not renormalized,
but the conserved current operator now has a negative anomalous 
dimension leading to an unwanted enhancement of $B\mu$.

This is why we added the coupling $\kappa (Q\overline Q)^2$ to the
superpotential in \eq{w-hidden}. The CFT without this coupling
has an exact non-anomalous $U(1)$ symmetry under which $X$ carries
charge 2, $Q$ and $\overline Q$ carry charge -1 and
$P$ and $\overline P$ carry charge 1. This symmetry is broken
by the messenger sector and therefore a coupling of the corresponding
current $2X^\dag X - Q^\dag Q - \overline Q^\dag \overline Q
+ P^\dag P + \overline P^\dag \overline P$ is generated by the
messenger interactions. Since the current is conserved in the CFT, 
it does not scale away. To fix the problem we must break the
$U(1)$ symmetry by strong CFT interactions. This is achieved with
the coupling $(Q\overline Q)^2$ which is relevant for $F < N$.

Having broken all dangerous $U(1)$ symmetries of the CFT we
know that all three of $k_X, k_Q, k_P$ renormalize strongly.
Unfortunately, there is no known technique for determining their
anomalous dimensions. Therefore we cannot determine
if this specific model does indeed sequester the $B_\mu$ operator. 
This is a general feature of models which employ our
mechanism: since anomalous dimensions must be large, they cannot
be computed in perturbation theory and even their sign is
unknown. Thus we cannot determine if any particular hidden sector model
suppresses $B_\mu$. But we expect that among the large number of
possible hidden sector CFTs there are some which have only positive anomalous
dimensions for the operators in question. In the Appendix we discuss
a simple perturbative example for a toy hidden sector (with no
supersymmetry breaking) where the anomalous dimensions
have the required signs for conformal sequestering of $B\mu$. 
We also show that conformal field theories with weakly
coupled Banks-Zaks fixed points do not have the correct signs for
all the relevant anomalous dimensions to be suitable as sequestering
hidden sectors.

We now discuss supersymmetry breaking.
At the scale $m$ (times renormalization factors
from the wave-functions of the fields) the hidden sector quarks obtain
masses and are integrated out of the theory. The remaining flavor-less
SQCD theory confines and generates a non-perturbative superpotential
which depends on the determinant of the masses of the quarks through
the matching of the holomorphic gauge coupling. The particular choice of 
masses and Yukawa couplings in \eq{w-hidden} ensures that this dynamical
superpotential does not depend on $X$. 
Therefore the low energy superpotential for $X$ only involves
the linear term $W=\Lambda^2 X$ and supersymmetry 
is broken spontaneously. The scalar expectation value for $X$ is stabilized
at the origin of field space by a non-trivial K\"ahler potential obtained from integrating out the 
hidden sector quarks.%
\footnote{A one loop diagram with quarks in the loop generates
the term $-(X^\dagger X)^2/m^2$ in the K\"ahler potential which stabilizes
the $X$ vev at the origin. However, the theory is strongly coupled
at the scale $m$ and we cannot be sure about the sign of the K\"ahler
term beyond the one-loop approximation.
A simple way to make this calculation reliable is to add $N$ 
new flavors of quarks $T + \overline T$ and an $N\times N$ matrix 
of singlets $Y$ with the superpotential 
$W_T=Y T\overline T -V^2 {\rm Tr} Y$.
This potential forces a complete breaking of the gauge symmetry 
at the scale $V \gg m$, all newly added fields and the gauge bosons
pick up masses of order $V$, and the low-energy theory of
$X, Q, \overline Q, P, \overline P$ is weakly coupled 
at the scale $m$ so that a perturbative calculation of the $X$
K\"ahler potential is reliable.
Note that we must reduce the number of flavors
of $Q$ and $P$ so that the theory remains a strongly coupled CFT
above the scale $V$ and the superpotential term
$(Q\overline Q)^2$ remains relevant. Using a-maximization
\cite{Intriligator:2003jj}
and checking the Seiberg dual of our theory we find
that the desired fixed point exists for $F$ near $N/2$.}

\begin{figure}[t]
\begin{center}
\includegraphics[width=0.7\textwidth]{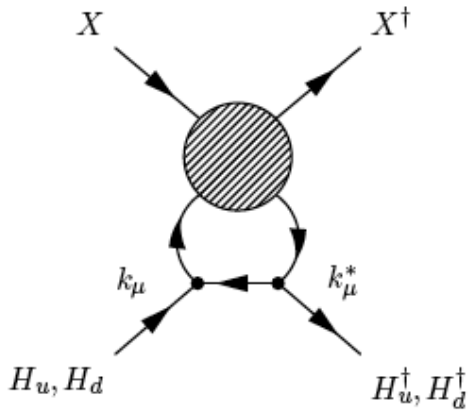}
\end{center}
\caption{The $\mu$-term renormalizes the Higgs soft masses.}
\label{fig:Higgs-RGE}
\end{figure}

What is the resulting pattern of MSSM soft masses? We obtain the
masses renormalized at the supersymmetry breaking scale by replacing
$X$ with it's expectation value $\langle \left. X \right|_F \rangle =F$.
Note that this expectation value is for the holomorphic field $X$, it is
not necessary to perform the wave function rescaling to switch to the
canonical $X$. Soft masses which stem from operators linear in
$X$ are then given by $ \sim \frac{g^2}{16 \pi^2} \frac{F}{M}$. 
All squark and slepton masses as well as $B\mu$ arise from
operators of the form $X^\dagger X$ times visible fields and are therefore
negligibly small at the supersymmetry breaking scale. 
They are regenerated by MSSM renormalization group running from the 
intermediate scale down to the weak scale. Except for the Higgs soft masses,
this spectrum is similar to the one of
gaugino mediation \cite{Kaplan:1999ac,Chacko:1999mi,Schmaltz:2000gy,
Cheng:2001an}
which has been studied in the literature \cite{Schmaltz:2000gy,Schmaltz:2000ei}.
Renormalization of the Higgs soft masses is different from the 
renormalization  of other soft masses. The difference is that the one-loop
diagram of \fig{Higgs-RGE} involving the $\mu$-operator of \eq{mu-op}
along with hidden sector interactions generates
the operator $ \int\! d^4\theta\,k_H X^\dag X H^\dag H/M^2$, where $H$ stands 
for either of the MSSM Higgses. The Higgs soft masses squared are then
expected to be of the same order as the $\mu$ term at the intermediate scale
because of this new additive contribution to their RGEs.
For details see 
\cite{Cohen:2006qc,crs2}.

We close with three comments about our mechanism for solving the $\mu/B_\mu$
problem. Firstly, our mechanism relies on strong hidden sector interactions
which are quite generic in theories of dynamical supersymmetry breaking.

Secondly, the strong suppression of scalar masses at the intermediate
scale also suppresses flavor-violation which may have entered the
scalar masses from high scale flavor physics.
Thus our mechanism for solving the $B_\mu$ problem
also makes the theory safer from FCNC constraints. Interestingly, 
this allows raising the messenger scale to near the GUT or string scale
without having to worry about flavor violation from stringy
physics.

Thirdly, throughout this work we used the holomorphic basis, which is convenient
because we are interested in the scaling of the $B_\mu$-term relative to
the $\mu$-term. The soft masses of MSSM superpartners are obtained
after replacing the holomorphic superfield $X$ with it's F-expectation 
value. In this basis the wave function renormalization factor $Z_X$
does not enter MSSM superpartner masses. However the $Z$-factor
does appear in the relationship between MSSM superpartner masses and
the gravitino mass \cite{Luty:2001jh,Dine:2004dv,Ibe:2005pj,Schmaltz:2006qs}.
Since we are relying on strong hidden sector interactions one expects 
that $Z_X \gg 1$ which makes the gravitino much heavier than the MSSM
superpartners.

Summary: conventional solutions to the $\mu/B_\mu$ problem in gauge mediation
rely either on complicated messenger sectors with tuned 
parameters and/or on extra light degrees of freedom. In this paper we 
propose a new solution. Our messenger sector is the simplest that 
generates $\mu$ of the same size as gaugino masses.
Assuming a positive sign for the relevant anomalous dimensions, the 
strong hidden sector interactions suppress the $B_\mu$ operator
such that $B_\mu \lesssim \mu^2$ at the intermediate scale where
the hidden sector interactions end.
MSSM interactions below the intermediate scale regenerate $B_\mu$.
We presented an explicit example for a hidden sector model which
realizes our mechanism. 

{\bf Note added:} During completion of the manuscript we learned that
Murayama, Nomura and Poland are developing a similar solution to the 
$\mu/B_\mu$ problem \cite{Murayama:2007ge}.

\begin{acknowledgments}
  This work was supported in part by the Department of Energy
  under grant no. DE-FG02-01ER-40676, grant no. DE-FG02-91ER-40676 and
  an Alfred P. Sloan Research Fellowship. MS acknowledges the support of
  the Aspen Center for Physics.
\end{acknowledgments} 

\section{Appendix}

In this appendix we discuss two examples of perturbative field
theories as toy models for the hidden sector and compute the anomalous
dimensions which determine whether operators of the form
$\frac{1}{M^2} \,k_{B}\, X^\dagger X H_u H_d$ sequester relative to
$\frac{1}{M}\,k_\mu\, X^\dagger H_u H_d$.

Our first toy hidden sector has only a single chiral superfield $X$
with the superpotential
\begin{equation}
W= \frac{\lambda}{3!} X^3 \ .
\end{equation}
In the holomorphic basis $k_\mu$ is not renormalized but
\begin{equation}
\frac{d}{dt} k_{B} = \gamma \, k_{B} = \frac{2 | \lambda |^2}{16 \pi^2}
\; k_{B} \ .
\end{equation}
Thus the anomalous dimension of $k_{B}$ is positive as desired, and
$B_\mu$ is sequestered relative to $\mu$. Of course, this toy hidden
sector is not suitable for our mechanism because
{\it ii.} Yukawa theories are not strongly coupled over a range of energy scales,
thus any sequestering effects are necessarily small and
{\it i.} it is too simple to include supersymmetry breaking.

As our second example we consider a conformal theory with a perturbative
Banks-Zaks (BZ) fixed point. We will show that BZ fixed point theories
necessarily have operators of the form
$\frac{1}{M^2} \,k_{B}\, X^\dagger X H_u H_d$
which do not sequester in the IR. Our proof of this statement will
depend on the fact that in BZ fixed point theories anomalous dimensions
are small so that
{\it i.} scaling dimensions of operators are close to free field values and
{\it ii.} one-loop anomalous dimensions dominate over higher loops.
Neither of these properties apply to the case of strongly coupled
CFTs which are needed to generate significant sequestering.
Therefore our ``no-go" result for BZ theories does not extend to
the theories of interest, but BZ theories provide a nice laboratory
to study the general mechanism of conformal sequestering.
\begin{figure}[t]
\begin{center}
\includegraphics[width=.9\textwidth]{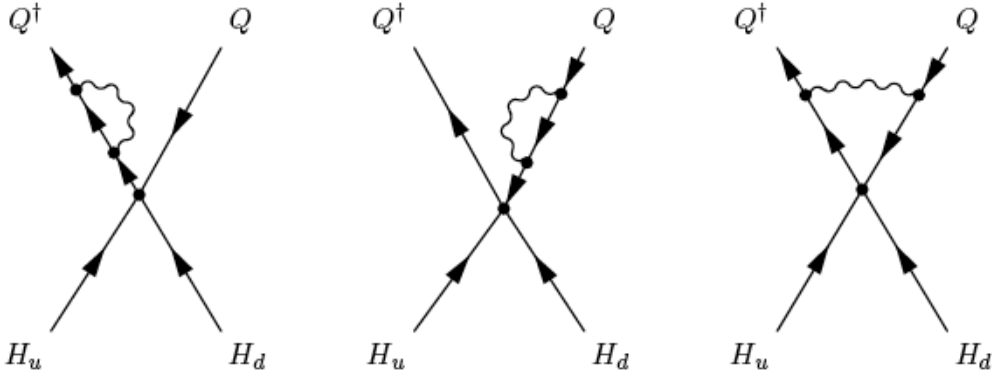}
\end{center}
\caption{Operators quadratic in hidden sector fields receive no corrections 
  due to hidden sector gauge interactions at one loop. Current conservation
  imply that the diagrams above cancel.}
\label{fig:A1}
\end{figure}

To start, we note that our BZ theory must contain a gauge invariant
fundamental chiral superfield $X$ so that the $\mu$-term operator
$\frac{1}{M}\,k_\mu\, X^\dagger H_u H_d$
is invariant and not suppressed by higher powers of the messenger scale $M$.
Since we want operators involving $X^\dagger X$ to sequester, $X$ must
interact. Then its scaling dimension $D[X]$ is greater than 1
by unitarity (in the canonical basis). We
adopt the canonical basis for the arguments of this example.
At weak coupling in a conformal BZ theory all marginal superpotential
operators are tri-linear in the fields. Hence the only possibility for
coupling $X$ to the conformal dynamics is to introduce a superpotential
$X Q \overline Q$ where $Q$ and $\overline Q$ stand for any (not necessarily 
distinct) chiral superfields which are charged under the
BZ gauge group.
This coupling gives a positive anomalous dimension to $X$ already at one loop. 
Therefore operators of the form $X^\dagger X$ must obtain even
larger positive anomalous dimensions at one loop in order for them to sequester.
We will now show that at least one operator involving $X^\dagger X$
has a vanishing anomalous dimension at one loop.

First note that in the canonical basis operators
quadratic in hidden sector fields ({\it e.g.} the $B_\mu$ operator)
do not receive corrections from hidden sector gauge interactions at one loop.
Current conservation at one loop
implies that the diagrams in \fig{A1} cancel with each other.
We therefore only need to focus on the renormalization due to Yukawa
couplings.

If $X$ is charged under a global $U(1)$ symmetry which
is unbroken by the Yukawa couplings, then the corresponding
current is protected by a non-renormalization theorem.
This current will involve the operator $X^\dagger X$ and
therefore the anomalous dimension of the $B\mu$ operator vanishes
(at one loop). Thus for sequestering to work
we must introduce Yukawa couplings to break any
global $U(1)$ symmetry under which $X$ is charged.
Absence of a global $U(1)$ symmetry implies that the
conformal R-symmetry which determines the scaling dimension
of $X$ must be uniquely determined from the superpotential
of the CFT. But since the superpotential only contains
tri-linear terms the only possible solution is that the
scaling dimensions of the fields in these Yukawa couplings are
all equal to 1. But then $X$ is a free field by unitarity
in contradiction to our initial assumptions.

We conclude that for weakly coupled theories we have a choice:
if we insist on conformal symmetry we are forced into allowing
global $U(1)$ symmetries of the superpotential leading to no sequestering (at one 
loop). Alternatively, we can give up conformal symmetry as in our first
example.

\begin{figure}[t]
\begin{center}
\includegraphics[width=.4\textwidth]{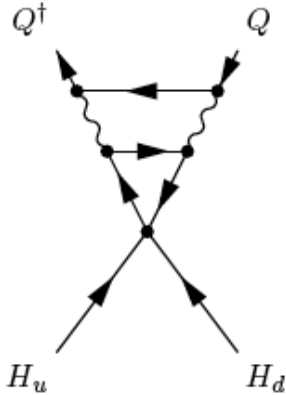}
\end{center}
\caption{Operators quadratic in hidden sector fields receive corrections 
  due to hidden sector gauge interactions at two loops.}
\label{fig:A2}
\end{figure}

Let us reiterate that this argument does not carry over to strongly
interacting CFTs for two reasons. One is that for strongly coupled
CFTs anomalous dimensions are large and superpotential couplings
which are higher order in fields may be marginal. Furthermore, gauge
interactions are now important to the renormalization. This is
because the cancellation in the renormalization of classically conserved
currents (\fig{A1}) does not extend to higher loops when there
are anomalies. For example, diagrams of the form shown in
\fig{A2}. do lead to sequestering of the bilinear $X^\dagger X$.
It was shown in \cite{Schmaltz:2006qs}
that in strongly coupled IR attractive conformal field theories with no
conserved global currents all operators of the form $X^\dagger X$ times
MSSM fields
have positive anomalous dimensions. For the purposes of solving the $\mu/B\mu$
problem as advocated in this paper we must further demand that these
anomalous dimensions are greater than twice the anomalous dimension of $X$.

\bibliography{reference}

\begin{thebibliography}{34}
\expandafter\ifx\csname natexlab\endcsname\relax\def\natexlab#1{#1}\fi
\expandafter\ifx\csname bibnamefont\endcsname\relax
  \def\bibnamefont#1{#1}\fi
\expandafter\ifx\csname bibfnamefont\endcsname\relax
  \def\bibfnamefont#1{#1}\fi
\expandafter\ifx\csname citenamefont\endcsname\relax
  \def\citenamefont#1{#1}\fi
\expandafter\ifx\csname url\endcsname\relax
  \def\url#1{\texttt{#1}}\fi
\expandafter\ifx\csname urlprefix\endcsname\relax\def\urlprefix{URL }\fi
\providecommand{\bibinfo}[2]{#2}
\providecommand{\eprint}[2][]{\url{#2}}

\bibitem[{\citenamefont{Dine et~al.}(1981)\citenamefont{Dine, Fischler, and
  Srednicki}}]{Dine:1981za}
\bibinfo{author}{\bibfnamefont{M.}~\bibnamefont{Dine}},
  \bibinfo{author}{\bibfnamefont{W.}~\bibnamefont{Fischler}}, \bibnamefont{and}
  \bibinfo{author}{\bibfnamefont{M.}~\bibnamefont{Srednicki}},
  \bibinfo{journal}{Nucl. Phys.} \textbf{\bibinfo{volume}{B189}},
  \bibinfo{pages}{575} (\bibinfo{year}{1981}).

\bibitem[{\citenamefont{Dimopoulos and Raby}(1981)}]{Dimopoulos:1981au}
\bibinfo{author}{\bibfnamefont{S.}~\bibnamefont{Dimopoulos}} \bibnamefont{and}
  \bibinfo{author}{\bibfnamefont{S.}~\bibnamefont{Raby}},
  \bibinfo{journal}{Nucl. Phys.} \textbf{\bibinfo{volume}{B192}},
  \bibinfo{pages}{353} (\bibinfo{year}{1981}).

\bibitem[{\citenamefont{Dine and Fischler}(1982)}]{Dine:1981gu}
\bibinfo{author}{\bibfnamefont{M.}~\bibnamefont{Dine}} \bibnamefont{and}
  \bibinfo{author}{\bibfnamefont{W.}~\bibnamefont{Fischler}},
  \bibinfo{journal}{Phys. Lett.} \textbf{\bibinfo{volume}{B110}},
  \bibinfo{pages}{227} (\bibinfo{year}{1982}).

\bibitem[{\citenamefont{Nappi and Ovrut}(1982)}]{Nappi:1982hm}
\bibinfo{author}{\bibfnamefont{C.~R.} \bibnamefont{Nappi}} \bibnamefont{and}
  \bibinfo{author}{\bibfnamefont{B.~A.} \bibnamefont{Ovrut}},
  \bibinfo{journal}{Phys. Lett.} \textbf{\bibinfo{volume}{B113}},
  \bibinfo{pages}{175} (\bibinfo{year}{1982}).

\bibitem[{\citenamefont{Alvarez-Gaume et~al.}(1982)\citenamefont{Alvarez-Gaume,
  Claudson, and Wise}}]{AlvarezGaume:1981wy}
\bibinfo{author}{\bibfnamefont{L.}~\bibnamefont{Alvarez-Gaume}},
  \bibinfo{author}{\bibfnamefont{M.}~\bibnamefont{Claudson}}, \bibnamefont{and}
  \bibinfo{author}{\bibfnamefont{M.~B.} \bibnamefont{Wise}},
  \bibinfo{journal}{Nucl. Phys.} \textbf{\bibinfo{volume}{B207}},
  \bibinfo{pages}{96} (\bibinfo{year}{1982}).

\bibitem[{\citenamefont{Dimopoulos and Raby}(1983)}]{Dimopoulos:1982gm}
\bibinfo{author}{\bibfnamefont{S.}~\bibnamefont{Dimopoulos}} \bibnamefont{and}
  \bibinfo{author}{\bibfnamefont{S.}~\bibnamefont{Raby}},
  \bibinfo{journal}{Nucl. Phys.} \textbf{\bibinfo{volume}{B219}},
  \bibinfo{pages}{479} (\bibinfo{year}{1983}).

\bibitem[{\citenamefont{Dine and Nelson}(1993)}]{Dine:1993yw}
\bibinfo{author}{\bibfnamefont{M.}~\bibnamefont{Dine}} \bibnamefont{and}
  \bibinfo{author}{\bibfnamefont{A.~E.} \bibnamefont{Nelson}},
  \bibinfo{journal}{Phys. Rev.} \textbf{\bibinfo{volume}{D48}},
  \bibinfo{pages}{1277} (\bibinfo{year}{1993}), \eprint{hep-ph/9303230}.

\bibitem[{\citenamefont{Dine et~al.}(1995)\citenamefont{Dine, Nelson, and
  Shirman}}]{Dine:1994vc}
\bibinfo{author}{\bibfnamefont{M.}~\bibnamefont{Dine}},
  \bibinfo{author}{\bibfnamefont{A.~E.} \bibnamefont{Nelson}},
  \bibnamefont{and} \bibinfo{author}{\bibfnamefont{Y.}~\bibnamefont{Shirman}},
  \bibinfo{journal}{Phys. Rev.} \textbf{\bibinfo{volume}{D51}},
  \bibinfo{pages}{1362} (\bibinfo{year}{1995}), \eprint{hep-ph/9408384}.

\bibitem[{\citenamefont{Dine et~al.}(1996)\citenamefont{Dine, Nelson, Nir, and
  Shirman}}]{Dine:1995ag}
\bibinfo{author}{\bibfnamefont{M.}~\bibnamefont{Dine}},
  \bibinfo{author}{\bibfnamefont{A.~E.} \bibnamefont{Nelson}},
  \bibinfo{author}{\bibfnamefont{Y.}~\bibnamefont{Nir}}, \bibnamefont{and}
  \bibinfo{author}{\bibfnamefont{Y.}~\bibnamefont{Shirman}},
  \bibinfo{journal}{Phys. Rev.} \textbf{\bibinfo{volume}{D53}},
  \bibinfo{pages}{2658} (\bibinfo{year}{1996}), \eprint{hep-ph/9507378}.

\bibitem[{\citenamefont{Giudice and Rattazzi}(1999)}]{Giudice:1998bp}
\bibinfo{author}{\bibfnamefont{G.~F.} \bibnamefont{Giudice}} \bibnamefont{and}
  \bibinfo{author}{\bibfnamefont{R.}~\bibnamefont{Rattazzi}},
  \bibinfo{journal}{Phys. Rept.} \textbf{\bibinfo{volume}{322}},
  \bibinfo{pages}{419} (\bibinfo{year}{1999}), \eprint{hep-ph/9801271}.

\bibitem[{\citenamefont{Dine and Mason}(2006)}]{Dine:2006xt}
\bibinfo{author}{\bibfnamefont{M.}~\bibnamefont{Dine}} \bibnamefont{and}
  \bibinfo{author}{\bibfnamefont{J.}~\bibnamefont{Mason}}
  (\bibinfo{year}{2006}), \eprint{hep-ph/0611312}.

\bibitem[{\citenamefont{Kitano et~al.}(2007)\citenamefont{Kitano, Ooguri, and
  Ookouchi}}]{Kitano:2006xg}
\bibinfo{author}{\bibfnamefont{R.}~\bibnamefont{Kitano}},
  \bibinfo{author}{\bibfnamefont{H.}~\bibnamefont{Ooguri}}, \bibnamefont{and}
  \bibinfo{author}{\bibfnamefont{Y.}~\bibnamefont{Ookouchi}},
  \bibinfo{journal}{Phys. Rev.} \textbf{\bibinfo{volume}{D75}},
  \bibinfo{pages}{045022} (\bibinfo{year}{2007}), \eprint{hep-ph/0612139}.

\bibitem[{\citenamefont{Murayama and
  Nomura}(2007{\natexlab{a}})}]{Murayama:2006yf}
\bibinfo{author}{\bibfnamefont{H.}~\bibnamefont{Murayama}} \bibnamefont{and}
  \bibinfo{author}{\bibfnamefont{Y.}~\bibnamefont{Nomura}},
  \bibinfo{journal}{Phys. Rev. Lett.} \textbf{\bibinfo{volume}{98}},
  \bibinfo{pages}{151803} (\bibinfo{year}{2007}{\natexlab{a}}),
  \eprint{hep-ph/0612186}.

\bibitem[{\citenamefont{Csaki et~al.}(2007)\citenamefont{Csaki, Shirman, and
  Terning}}]{Csaki:2006wi}
\bibinfo{author}{\bibfnamefont{C.}~\bibnamefont{Csaki}},
  \bibinfo{author}{\bibfnamefont{Y.}~\bibnamefont{Shirman}}, \bibnamefont{and}
  \bibinfo{author}{\bibfnamefont{J.}~\bibnamefont{Terning}},
  \bibinfo{journal}{JHEP} \textbf{\bibinfo{volume}{05}}, \bibinfo{pages}{099}
  (\bibinfo{year}{2007}), \eprint{hep-ph/0612241}.

\bibitem[{\citenamefont{Aharony and Seiberg}(2007)}]{Aharony:2006my}
\bibinfo{author}{\bibfnamefont{O.}~\bibnamefont{Aharony}} \bibnamefont{and}
  \bibinfo{author}{\bibfnamefont{N.}~\bibnamefont{Seiberg}},
  \bibinfo{journal}{JHEP} \textbf{\bibinfo{volume}{02}}, \bibinfo{pages}{054}
  (\bibinfo{year}{2007}), \eprint{hep-ph/0612308}.

\bibitem[{\citenamefont{Murayama and
  Nomura}(2007{\natexlab{b}})}]{Murayama:2007fe}
\bibinfo{author}{\bibfnamefont{H.}~\bibnamefont{Murayama}} \bibnamefont{and}
  \bibinfo{author}{\bibfnamefont{Y.}~\bibnamefont{Nomura}},
  \bibinfo{journal}{Phys. Rev.} \textbf{\bibinfo{volume}{D75}},
  \bibinfo{pages}{095011} (\bibinfo{year}{2007}{\natexlab{b}}),
  \eprint{hep-ph/0701231}.

\bibitem[{\citenamefont{Delgado et~al.}(2007)\citenamefont{Delgado, Giudice,
  and Slavich}}]{Delgado:2007rz}
\bibinfo{author}{\bibfnamefont{A.}~\bibnamefont{Delgado}},
  \bibinfo{author}{\bibfnamefont{G.~F.} \bibnamefont{Giudice}},
  \bibnamefont{and} \bibinfo{author}{\bibfnamefont{P.}~\bibnamefont{Slavich}}
  (\bibinfo{year}{2007}), \eprint{arXiv:0706.3873 [hep-ph]}.

\bibitem[{\citenamefont{Abel et~al.}(2007)\citenamefont{Abel, Durnford,
  Jaeckel, and Khoze}}]{Abel:2007jx}
\bibinfo{author}{\bibfnamefont{S.}~\bibnamefont{Abel}},
  \bibinfo{author}{\bibfnamefont{C.}~\bibnamefont{Durnford}},
  \bibinfo{author}{\bibfnamefont{J.}~\bibnamefont{Jaeckel}}, \bibnamefont{and}
  \bibinfo{author}{\bibfnamefont{V.~V.} \bibnamefont{Khoze}}
  (\bibinfo{year}{2007}), \eprint{arXiv:0707.2958 [hep-ph]}.

\bibitem[{\citenamefont{Dvali et~al.}(1996)\citenamefont{Dvali, Giudice, and
  Pomarol}}]{Dvali:1996cu}
\bibinfo{author}{\bibfnamefont{G.~R.} \bibnamefont{Dvali}},
  \bibinfo{author}{\bibfnamefont{G.~F.} \bibnamefont{Giudice}},
  \bibnamefont{and} \bibinfo{author}{\bibfnamefont{A.}~\bibnamefont{Pomarol}},
  \bibinfo{journal}{Nucl. Phys.} \textbf{\bibinfo{volume}{B478}},
  \bibinfo{pages}{31} (\bibinfo{year}{1996}), \eprint{hep-ph/9603238}.

\bibitem[{\citenamefont{Giudice and Masiero}(1988)}]{Giudice:1988yz}
\bibinfo{author}{\bibfnamefont{G.~F.} \bibnamefont{Giudice}} \bibnamefont{and}
  \bibinfo{author}{\bibfnamefont{A.}~\bibnamefont{Masiero}},
  \bibinfo{journal}{Phys. Lett.} \textbf{\bibinfo{volume}{B206}},
  \bibinfo{pages}{480} (\bibinfo{year}{1988}).

\bibitem[{\citenamefont{Dine et~al.}(2004)}]{Dine:2004dv}
\bibinfo{author}{\bibfnamefont{M.}~\bibnamefont{Dine}} \bibnamefont{et~al.},
  \bibinfo{journal}{Phys. Rev.} \textbf{\bibinfo{volume}{D70}},
  \bibinfo{pages}{045023} (\bibinfo{year}{2004}), \eprint{hep-ph/0405159}.

\bibitem[{\citenamefont{Cohen et~al.}(2007)\citenamefont{Cohen, Roy, and
  Schmaltz}}]{Cohen:2006qc}
\bibinfo{author}{\bibfnamefont{A.~G.} \bibnamefont{Cohen}},
  \bibinfo{author}{\bibfnamefont{T.~S.} \bibnamefont{Roy}}, \bibnamefont{and}
  \bibinfo{author}{\bibfnamefont{M.}~\bibnamefont{Schmaltz}},
  \bibinfo{journal}{JHEP} \textbf{\bibinfo{volume}{02}}, \bibinfo{pages}{027}
  (\bibinfo{year}{2007}), \eprint{hep-ph/0612100}.

\bibitem[{\citenamefont{Cohen et~al.}()\citenamefont{Cohen, Roy, and
  Schmaltz}}]{crs2}
\bibinfo{author}{\bibfnamefont{A.~G.} \bibnamefont{Cohen}},
  \bibinfo{author}{\bibfnamefont{T.}~\bibnamefont{Roy}}, \bibnamefont{and}
  \bibinfo{author}{\bibfnamefont{M.}~\bibnamefont{Schmaltz}} (????),
  \bibinfo{note}{to appear}.

\bibitem[{\citenamefont{Seiberg}(1995)}]{Seiberg:1994pq}
\bibinfo{author}{\bibfnamefont{N.}~\bibnamefont{Seiberg}},
  \bibinfo{journal}{Nucl. Phys.} \textbf{\bibinfo{volume}{B435}},
  \bibinfo{pages}{129} (\bibinfo{year}{1995}), \eprint{hep-th/9411149}.

\bibitem[{\citenamefont{Schmaltz and Sundrum}(2006)}]{Schmaltz:2006qs}
\bibinfo{author}{\bibfnamefont{M.}~\bibnamefont{Schmaltz}} \bibnamefont{and}
  \bibinfo{author}{\bibfnamefont{R.}~\bibnamefont{Sundrum}},
  \bibinfo{journal}{JHEP} \textbf{\bibinfo{volume}{11}}, \bibinfo{pages}{011}
  (\bibinfo{year}{2006}), \eprint{hep-th/0608051}.

\bibitem[{\citenamefont{Intriligator and Wecht}(2003)}]{Intriligator:2003jj}
\bibinfo{author}{\bibfnamefont{K.}~\bibnamefont{Intriligator}}
  \bibnamefont{and} \bibinfo{author}{\bibfnamefont{B.}~\bibnamefont{Wecht}},
  \bibinfo{journal}{Nucl. Phys.} \textbf{\bibinfo{volume}{B667}},
  \bibinfo{pages}{183} (\bibinfo{year}{2003}), \eprint{hep-th/0304128}.

\bibitem[{\citenamefont{Kaplan et~al.}(2000)\citenamefont{Kaplan, Kribs, and
  Schmaltz}}]{Kaplan:1999ac}
\bibinfo{author}{\bibfnamefont{D.~E.} \bibnamefont{Kaplan}},
  \bibinfo{author}{\bibfnamefont{G.~D.} \bibnamefont{Kribs}}, \bibnamefont{and}
  \bibinfo{author}{\bibfnamefont{M.}~\bibnamefont{Schmaltz}},
  \bibinfo{journal}{Phys. Rev.} \textbf{\bibinfo{volume}{D62}},
  \bibinfo{pages}{035010} (\bibinfo{year}{2000}), \eprint{hep-ph/9911293}.

\bibitem[{\citenamefont{Chacko et~al.}(2000)\citenamefont{Chacko, Luty, Nelson,
  and Ponton}}]{Chacko:1999mi}
\bibinfo{author}{\bibfnamefont{Z.}~\bibnamefont{Chacko}},
  \bibinfo{author}{\bibfnamefont{M.~A.} \bibnamefont{Luty}},
  \bibinfo{author}{\bibfnamefont{A.~E.} \bibnamefont{Nelson}},
  \bibnamefont{and} \bibinfo{author}{\bibfnamefont{E.}~\bibnamefont{Ponton}},
  \bibinfo{journal}{JHEP} \textbf{\bibinfo{volume}{01}}, \bibinfo{pages}{003}
  (\bibinfo{year}{2000}), \eprint{hep-ph/9911323}.

\bibitem[{\citenamefont{Schmaltz and
  Skiba}(2000{\natexlab{a}})}]{Schmaltz:2000gy}
\bibinfo{author}{\bibfnamefont{M.}~\bibnamefont{Schmaltz}} \bibnamefont{and}
  \bibinfo{author}{\bibfnamefont{W.}~\bibnamefont{Skiba}},
  \bibinfo{journal}{Phys. Rev.} \textbf{\bibinfo{volume}{D62}},
  \bibinfo{pages}{095005} (\bibinfo{year}{2000}{\natexlab{a}}),
  \eprint{hep-ph/0001172}.

\bibitem[{\citenamefont{Cheng et~al.}(2001)\citenamefont{Cheng, Kaplan,
  Schmaltz, and Skiba}}]{Cheng:2001an}
\bibinfo{author}{\bibfnamefont{H.~C.} \bibnamefont{Cheng}},
  \bibinfo{author}{\bibfnamefont{D.~E.} \bibnamefont{Kaplan}},
  \bibinfo{author}{\bibfnamefont{M.}~\bibnamefont{Schmaltz}}, \bibnamefont{and}
  \bibinfo{author}{\bibfnamefont{W.}~\bibnamefont{Skiba}},
  \bibinfo{journal}{Phys. Lett.} \textbf{\bibinfo{volume}{B515}},
  \bibinfo{pages}{395} (\bibinfo{year}{2001}), \eprint{hep-ph/0106098}.

\bibitem[{\citenamefont{Schmaltz and
  Skiba}(2000{\natexlab{b}})}]{Schmaltz:2000ei}
\bibinfo{author}{\bibfnamefont{M.}~\bibnamefont{Schmaltz}} \bibnamefont{and}
  \bibinfo{author}{\bibfnamefont{W.}~\bibnamefont{Skiba}},
  \bibinfo{journal}{Phys. Rev.} \textbf{\bibinfo{volume}{D62}},
  \bibinfo{pages}{095004} (\bibinfo{year}{2000}{\natexlab{b}}),
  \eprint{hep-ph/0004210}.

\bibitem[{\citenamefont{Luty and Sundrum}(2002)}]{Luty:2001jh}
\bibinfo{author}{\bibfnamefont{M.~A.} \bibnamefont{Luty}} \bibnamefont{and}
  \bibinfo{author}{\bibfnamefont{R.}~\bibnamefont{Sundrum}},
  \bibinfo{journal}{Phys. Rev.} \textbf{\bibinfo{volume}{D65}},
  \bibinfo{pages}{066004} (\bibinfo{year}{2002}), \eprint{hep-th/0105137}.

\bibitem[{\citenamefont{Ibe et~al.}(2006)\citenamefont{Ibe, Izawa, Nakayama,
  Shinbara, and Yanagida}}]{Ibe:2005pj}
\bibinfo{author}{\bibfnamefont{M.}~\bibnamefont{Ibe}},
  \bibinfo{author}{\bibfnamefont{K.~I.} \bibnamefont{Izawa}},
  \bibinfo{author}{\bibfnamefont{Y.}~\bibnamefont{Nakayama}},
  \bibinfo{author}{\bibfnamefont{Y.}~\bibnamefont{Shinbara}}, \bibnamefont{and}
  \bibinfo{author}{\bibfnamefont{T.}~\bibnamefont{Yanagida}},
  \bibinfo{journal}{Phys. Rev.} \textbf{\bibinfo{volume}{D73}},
  \bibinfo{pages}{015004} (\bibinfo{year}{2006}), \eprint{hep-ph/0506023}.

\bibitem[{\citenamefont{Murayama et~al.}(2008)\citenamefont{Murayama, Nomura,
  and Poland}}]{Murayama:2007ge}
\bibinfo{author}{\bibfnamefont{H.}~\bibnamefont{Murayama}},
  \bibinfo{author}{\bibfnamefont{Y.}~\bibnamefont{Nomura}}, \bibnamefont{and}
  \bibinfo{author}{\bibfnamefont{D.}~\bibnamefont{Poland}},
  \bibinfo{journal}{Phys. Rev.} \textbf{\bibinfo{volume}{D77}},
  \bibinfo{pages}{015005} (\bibinfo{year}{2008}), \eprint{arXiv:0709.0775
  [hep-ph]}.

\end{thebibliography}

\end{document}